\documentstyle[preprint,eqsecnum,aps]{revtex}

\begin{document}

\title {BRST operator quantization of generally covariant gauge systems}

\author{Rafael Ferraro{\footnote{Electronic address:  ferraro@iafe.uba.ar}}}

\address{{\tighten
{\it Instituto de Astronom\'\i a y F\'\i sica del Espacio,\\ Casilla de
Correo 67 - Sucursal 28,\\1428 Buenos Aires, Argentina\\ and Departamento
de F\'\i sica, Facultad de Ciencias Exactas y Naturales,\\ Universidad de
Buenos Aires - Ciudad Universitaria, Pabell\' on I\\ 1428 Buenos Aires,
Argentina\\}}}

\author{Daniel M. Sforza{\footnote{Electronic address: sforza@iafe.uba.ar}}}

\address{{\tighten
{\it Instituto de Astronom\'\i a y F\'\i sica del Espacio,\\
Casilla de Correo 67 - Sucursal 28,\\1428 Buenos Aires, Argentina\\}}}

\maketitle

\begin{abstract}

{\tighten

 The BRST generator is realized as a Hermitian nilpotent operator for a
finite-dimensional gauge system featuring a quadratic super-Hamiltonian
and linear supermomentum constraints.  As a
result, the emerging ordering for the Hamiltonian constraint is not trivial,
because the potential must enter the kinetic term in order to obtain a
quantization invariant under scaling. Namely, BRST
quantization does not lead to the curvature term used in the literature as
a means to get that invariance. The inclusion of the potential in the
kinetic term, far from being unnatural, is beautifully justified in light of
the Jacobi's principle.

}

\end{abstract}  
\vskip  1cm 


\narrowtext

\newpage

\section{Introduction}

The gravitational field is a generally covariant system with a Hamiltonian
which is constrained to vanish.  Actually, the ``geometrodynamical''
Hamiltonian is a linear combination of four constraints (in each space
point); three of them are the supermomenta (linear and homogeneous
functions of the field momenta), and the other one is the super-Hamiltonian (a
quadratic function in the field momenta).  The quantization of such a system
requires searching for a factor ordering that leads to constraint operators
preserving the algebra of constraints (absence of anomalies).  This issue
is unsolved yet \cite{v1,v2}.  In order to deal with a more simple system,
it is common to freeze most degrees of freedom to end with a
finite-dimensional system, featuring constraints that resemble those of the
gravitational field ({\it minisuperspace} models). In this spirit,
H\'aj\'{\i}\v{c}ek and Kucha\v{r} \cite{hk90} have studied the quantization
of one such finite dimensional system in the context of Dirac quantization.

In addition to the Dirac method, the Becci-Rouet-Stora-Tyutin (BRST)
formalism  is a powerful tool to quantize
a first class constrained system.  BRST method is based on the realization
of the generator of a rigid supersymmetry, the BRST symmetry, as a
Hermitian nilpotent operator, the physical quantum states being picked up
from the cohomology of this operator. The power of the BRST formalism
consists in the automatic invariance of the quantization under combinations
of constraints, because these combinations are equivalent to coordinate
changes in the fermionic sector.

Our aim is to perform the BRST quantization of a system such us the one
studied in Ref. [3], i.e.,  a generally covariant system described by $n$
canonical pairs $(q^i,p_i)$ subjected to $m+1$ first class constraints,
where $m$ of them are linear and homogeneous in the momenta, and the other
one is a quadratic function of the momenta, with an indefinite nondegenerate
metric $G^{ij}$, plus a nonvanishing potential $V$.

The paper is organized as follows. In Sec. II we review the BRST
quantization for a system of linear constraints. It is emphasized that the
usual ghost contribution to the constraint operator (the anti-Hermitian
term ${i\over 2} C^b_{ab}$) is related to the  volume density 
induced by the constraints on the gauge orbit.  In Sec. III we add a
quadratic constraint.  We take advantage of the possibility of factorizing
out the nonvanishing potential in the Hamiltonian constraint; this is
equivalent to scale the constraint for obtaining an equivalent one with a
constant potential.  So, we first get the Hermitian nilpotent BRST
generator $\hat\Omega$ for a constant potential, and then the original
potential is reentered by means of a unitary transformation.  The volume
density 
of Sec. II plays an essential and elegant role in guessing the
ordering that leads to the nilpotency. In Sec. IV we look for the
constraint operators of the Dirac method. In the BRST formalism they can
be got from $\hat\Omega$ after a suitable ordering of the ghost sector.  As
it happens with the linear constraints, the Hamiltonian constraint also gets
a non-Hermitian ghost contribution.  Section V is devoted to the conclusions.
We emphasize the  role  played  by  the ghost contributions to the constraint
operators in preserving the  invariance  of the theory under the relevant 
combinations of  constraints  (those  which  do  not  change  the  form of the
constraints).    We reduce the system by
fixing the gauge freedom coming from the linear constraints, and we show that
the ghost contribution to the super-Hamiltonian leads to  the  emerging of
the Laplacian associated with the scale-invariant metric  $V G_{ij}$,  in a
beautiful agreement with the Jacobi's principle.  Namely, no  curvature  term
is needed to get the invariance under scaling.

\section{Linear Constraints} 

For a system of $m$ linearly independent constraints
\begin{equation} G_a(q^i,p_j)=\xi^k_a(q^i)p_k, ~~~~~~~~~a=1,...,m,
\label{vle}\end{equation}
the problem of finding a factor ordering satisfying the algebra
\begin{equation}\{G_a,G_b\}=C_{ab}^c(q) G_c \label{cl}\end{equation}
is trivially solved by
\begin{equation}
\hat  G_a=  f^{1\over  2}    \xi^i_a   \hat  p_i  f^{-{1\over
2}},\label{vlinq}\end{equation} 
where $f$ is arbitrary.  In the Dirac quantization the function $f$ can be
determined by asking the constraint operators to preserve the geometrical
character of the wave function \cite{hk90,mp89,fhp}. This character is
determined by the transformation law of the wave function under the changes
that leave invariant the classical theory: coordinate changes and linear
combinations of constraints.  The wave function should change in such a way
that the physical inner product remains unchanged.

\bigskip
 
On the other hand, in the BRST formalism, the original phase space is
extended by including a canonically conjugate pair of ghost $(\eta^a,{\cal
P}_a)$ for each constraint, with  opposite parity.  The central object
is the BRST generator, a fermionic function
$\Omega=\Omega(q^i,p_j,\eta^a,{\cal P}_b)$ that captures all the
identities satisfied by the system of first class constraints in the
equation
\begin{equation}\{\Omega,\Omega\}=0.\label{nil}\end{equation}

The existence of $\Omega$ is guaranteed at the classical level, and
$\Omega$ is unique up to canonical transformations in the extended phase
space.  It can be built by means of a recursive method \cite{ht}.  The
result for the system (\ref{vle}) and (\ref{cl}) is
\begin{equation}
\Omega^{linear}=\eta^a  G_a  +    {1\over  2}  \eta^a  \eta^b
C_{ab}^c {\cal P}_c .\label{omegacl}\end{equation} 

In order to quantize the extended system, the classical BRST generator must
be realized as a Hermitian operator. The theory is free from BRST
anomalies, if a Hermitian realization of $\Omega$ can be found such that
the classical property (\ref{nil}) becomes
\begin{equation}
[{\hat\Omega},{\hat\Omega}]=2{\hat\Omega}^2=0 ,\label{nilq}\end{equation} 
i.e.,  ${\hat{\Omega}}$ must be nilpotent.  The BRST physical quantum states
belong to the set of equivalence classes of BRST-closed states
(${\hat{\Omega}}\psi=0$) moduli BRST-exact ones ($\psi={\hat{\Omega}}\chi$)
(quantum BRST cohomology).

Let us adopt the notation used in Ref.\ \onlinecite{fhp}:
\begin{equation} 
\eta^{c_s}=(q^i, \eta^a),~~~~~~~~~~~{\cal P}_{c_s}=(p_i,{\cal
P}_a),~~~~~~~~~~~~~~~~s=-1,0.\label{not}\end{equation} Then,
$\Omega^{linear}$ can be written as
\begin{equation}\Omega^{linear}=\sum_{s=-1}^0 \Omega^{c_s}{\cal P}_{c_s},
\label{omlin}\end{equation}  
where
\begin{equation}\Omega^{c_s}\equiv (\eta^a\xi^i_a~,~{1\over 2}\eta^a\eta^b
C_{ab}^c). \end{equation} 

The ordering
\begin{equation}
\hat{\Omega}^{linear}=\sum_{s=-1}^0f^{1\over   2}\Omega^{c_s}
\hat{\cal P}_{c_s}f^{-{1\over 2}}\label{ordlin}\end{equation} 
is nilpotent for any $f(q)$ [it is just the classical result (\ref{nil})]
but $f$ should be chosen in such a way that ${\hat\Omega}^{linear}$ is
Hermitian. It results that $f$ must satisfy
\begin{equation}C^b_{ab}=f^{-1}(f\xi^i_a)_{,i}.\label{div}\end{equation}    

The obtained $\hat{\Omega}^{linear}$ could be also obtained by symmetrizing
Eq.  (\ref{omlin}).  This realization of $\hat{\Omega}^{linear}$ leads to
constraint operators that coincide with the ones obtained in the
geometrical Dirac method (see, for example, Ref. \cite{fhp} and references
therein):
\begin{equation}
\hat G_a=  f^{1\over  2}    \xi^i_a   \hat  p_i  f^{-{1\over
2}}=\xi^i_a \hat p_i - {i\over 2} \xi^i_{a,i} + {i\over 2}
C^b_{ab}.\label{vlinqe}\end{equation}
Although Eq.  (\ref{div}) is all one needs to establish
$\hat{\Omega}^{linear}$, it does not univocally define $f$. In fact, the
right hand-side does not change if $f$ is multiplied by a gauge-invariant
function.  The following proposition will  make clear the geometrical meaning
of $f$ in Eq. (\ref{div}).

{\it Proposition.}
For a given set (\ref{vle}) and (\ref{cl}), let $\tilde\alpha$ be
a volume induced by the constraints in the original configuration
space $M$:
\begin{equation}
{\tilde\alpha}\equiv {\tilde  E}^1\wedge...\wedge   {\tilde  E}^m
\wedge {\tilde \omega},\label{volnat}\end{equation}
where $\{\tilde E^a\}$ is the dual basis of $\{\vec\xi_a\}$ in $T_{||}M$,
the  longitudinal  tangent  space;    and  ${\tilde  \omega}\  =\  \omega(y)\
dy^1\wedge...\wedge
dy^{n-m}$ is a closed $n-m$ form,  the $y^r$'s being $n-m$ functions
which are left invariant by the gauge transformations generated by the
linear constraints,\footnote{We do not call these functions ``observables''
because they will not be invariant under the action of the quadratic
constraint we are going to introduce later.} i.e.,
$dy^r(\vec\xi_a)=0~~\forall  r,a$.  $\tilde\alpha$ is the volume  induced  by
the  constraints  in the gauge orbit, times a  (nonchosen)  volume  in  the
``reduced" space. Then,
\begin{equation}
C_{ab}^b=div_{\tilde\alpha}~\vec\xi_a.\label{prop1}\end{equation}
{\it Proof.} We will take advantage of the fact that any basis can be
(locally) Abelianized.  So, we will prove the proposition for an Abelian
basis, and then we will transform  both sides of Eq.  (\ref{prop1})
showing that they remain equal for an arbitrary basis of $T_{||}M$.

Let $\{{\vec\xi}^{\prime}_a\}$ be an Abelian basis in $T_{||}M$,
then the left-hand side of Eq.  (\ref{prop1}) is $C^{\prime b}_{ab}=0$.  On
the other hand, the ${\tilde\alpha}^{\prime}$ divergence of a vector field
${\vec\xi}^{\prime}_a$ is written, by definition\cite{schutz}, in terms of
the exterior derivative of the $(n-1)$-form
${\tilde\alpha}^{\prime}({\vec\xi}^{\prime}_a)$:
\begin{equation}
(div_{{\tilde\alpha}^{\prime}}{\vec\xi}^{\prime}_a){\tilde\alpha}^{\prime}
\equiv d[{\tilde\alpha}^{\prime}({\vec\xi}^{\prime}_a)].
\label{defdiv}\end{equation}
The right-hand side of Eq.  (\ref{prop1}) is also zero because
${\tilde\alpha}^{\prime}({\vec\xi}^{\prime}_a)$ is closed. In
fact, the forms ${\tilde E}^{\prime a}$ are (locally) exact, since an
Abelian basis is a coordinate basis.  Then, Eq.  (\ref{prop1}) is proved
for Abelianized constraints.

Now, let us change the basis
\begin{equation}
{\vec\xi}_a=A_a^{~b}(q)~{\vec\xi}^{\prime}_b,~~~~~~~~~~~~{\tilde
E}^a=A^a_{~b}(q)~{\tilde E}^{\prime b}
\label{cbase}\end{equation}
($A^a_{~b}$ being the matrix inverse to $A_a^{~b}$). Then, 
\begin{equation} 
C^b_{ab}=E^b_i(\xi^j_b    \xi^i_{a,j}    -\xi^j_a  \xi^i_{b,j})=
A^b_{~c}(A^{~c}_{a,j}\xi^j_b - A^{~c}_{b,j}\xi^j_a).
\label{stf}\end{equation}

On the other hand,
\begin{eqnarray}
d[{\tilde\alpha}(\vec\xi_a)]&&=d[det
A^{-1}{\tilde\alpha}^{\prime}(\vec\xi_a)] =d[A_a^{~b}det
A^{-1}{\tilde\alpha}^{\prime}({\vec\xi}^{\prime}_a)]\nonumber\\
&&=\sum_{b=1}^{m} (-1)^{b-1}(A_a^{~b}~det A^{-1})_{,j}~ dq^j \wedge {\tilde
E}^{\prime 1}\wedge...\nonumber\\ &&~~~...\wedge {\tilde E}^{\prime b-1}
\wedge {\tilde E}^{\prime b+1}\wedge ...\wedge {\tilde E}^{\prime m} \wedge
{\tilde \omega}\nonumber\\ && =(A_a^{~b}~det
A^{-1})_{,j}~ A^c_{~b}~ \xi^j_c~ {\tilde E}^{\prime 1}\wedge...\wedge
{\tilde E}^{\prime m}\wedge {\tilde \omega},\end{eqnarray}
($det A^{-1}\equiv det A^a_{~b}$), because only the 
component $dq^j({\vec\xi}^{\prime}_b)={\xi}^{\prime j}_b= 
A^c_{~b}\xi^j_c$ contributes.

Therefore
\begin{equation}
d[{\tilde\alpha}(\vec\xi_a)]=A^b_{~c}(A^{~c}_{a,j}\xi^j_b-A^{~c}_{b,j}
\xi^j_a) {\tilde\alpha}.\label{stf'}\end{equation} 
Thus, Eqs. (\ref{stf}) and (\ref{stf'}) tell us that both sides of
Eq. (\ref{div}) have the same value whatever the basis of $T_{||}M$
is. Then, the proposition results to be true for any set of linear and
homogeneous first class constraints.

The  result of the proposition means that $f$ in  Eq.    (\ref{div})  can  be
regarded as the component of $\tilde\alpha$ in the coordinate basis $\{dq^i\}$:
\begin{equation}
{\tilde\alpha}= f~dq^1\wedge...\wedge dq^n.
\end{equation}

\bigskip

At the level of BRST, a redefinition of the constraints such as the one of Eq.
(\ref{cbase}) is regarded as a change of variables $\eta^a \rightarrow
\eta^{\prime a}=\eta^b A_b^{~a}(q)$.  Since the BRST wave function behaves
as a superdensity of weight 1/2 in the space ($q,~\eta$) (in order to
leave the inner product invariant), one concludes that the factors $f^{1
\over 2}$ and $f^{-{1 \over 2}}$ in Eq.  (\ref{ordlin}) are exactly what is
needed in order for ${\hat{\Omega}^{linear}}\psi$ to behave in the same way as
$\psi$ under such a change (in fact, $f \rightarrow f^{\prime}=det A~f$).
This property of $f$ should be taken into account at the moment of
quantizing a system with a quadratic constraint, because it could
facilitate the searching for the operator $\hat\Omega$.

\section{Quadratic Constraint}

As it was stated in the introduction, we are going to consider a quadratic
constraint with a nonvanishing potential.  This property enables us to
factorize out the potential, and replace the quadratic constraint by an
equivalent one with a constant potential.  So, let us begin by considering a
Hamiltonian constraint $h(q^i,p_j)$:
\begin{equation}  
h(q^k,p_j)={1\over 2} g^{ij}(q^k) p_i p_j+\lambda, ~~~~~~\lambda=const,
\label{vce}\end{equation} $g^{ij}$ being an indefinite nondegenerate metric.
A more general nonvanishing potential $V = \lambda \vartheta(q)$ will 
enter later.  

In order that the set of constraints remains first class, we demand [together
with the relations (\ref{cl})],
\begin{equation}\{h,G_a\}=c_{oa}^b(q,p) G_b, \label{cc}\end{equation}
where 
\begin{equation}c_{oa}^b(q,p)=c_{oa}^{bj}(q) p_j. \label{fec}\end{equation}

\bigskip

Since one has added a constraint, the already extended phase space must be
further extended by adding the pair $(\eta^o, {\cal P}_o)$ associated with
$h$.  One finds that the new BRST generator is
\begin{equation}
\Omega=\eta^o h +  \eta^a  G_a + \eta^o \eta^a c_{oa}^b {\cal
P}_b + {1\over 2} \eta^a \eta^b  C_{ab}^c  {\cal P}_c
\equiv\Omega^{quad}+\Omega^{linear},\label{omegacl2}\end{equation} 
where  $\Omega^{linear}$  is the one of Eq. (\ref{omlin}),  and
$\Omega^{quad}$ is 
\begin{equation}\Omega^{quad}={1\over  2}\sum_{r,s=-1}^0  {\cal  P}_{a_r}
\Omega^{a_rb_s} {\cal P}_{b_s}+
\eta^o\lambda,\label{omcuad}\end{equation} with
\begin{equation}
\Omega^{a_rb_s} \equiv\pmatrix{ \eta^o g^{ij} &\eta^o\eta^a
c_{oa}^{bi}\cr ~& ~\cr \eta^o\eta^b c_{ob}^{aj} &0\cr}. \end{equation} 

One quantizes the system by turning the BRST generator in a Hermitian and
nilpotent operator $\hat\Omega$:
\begin{equation}[\hat{{\Omega}},\hat{{\Omega}}]=[\hat\Omega^{quad},
\hat\Omega^{quad}]+2[\hat\Omega^{quad},\hat\Omega^{linear}]+
[\hat\Omega^{linear},\hat\Omega^{linear}]=0.\label{nilpo}\end{equation} 

The term $[\hat\Omega^{quad},\hat\Omega^{quad}]$ is zero trivially because
${\eta^o}^2=0$ (note that $\Omega$ does not depend on ${\cal P}_o$).  The
last term is zero because $\hat\Omega^{linear}$ is already nilpotent.  So,
we only must find an ordering for $\hat\Omega^{quad}$ satisfying
$[\hat\Omega^{quad},\hat\Omega^{linear}]=0$.  The structure of
$\hat{\Omega}^{linear}$ strongly suggests the following Hermitian ordering
for $\hat{\Omega}^{quad}$: 
\begin{equation}\hat{\Omega}^{quad}={1\over 2}
\sum_{r,s=-1}^0 f^{-{1\over  2}}\hat{\cal  P}_{a_r} f \Omega^{a_rb_s}
\hat{\cal P}_{b_s}f^{-{1\over 2}}+\eta^o\lambda. 
\label{ordcuad}\end{equation} 

In fact, it is proved by direct calculation that $\hat\Omega$ results to be
nilpotent.

\section{Constraint Operators}

In this section we are going to identify the Dirac constraint
operators. They can be easily found by casting the Hermitian and nilpotent
operator $\hat\Omega$, the sum of Eqs. (\ref{ordlin})-(\ref{ordcuad}), in the
appropiate form. As in  Sec. 14.5 of Ref.\cite{ht}, we define the
constraint operators of the Dirac method to be the coefficients of the
ghost operators in the BRST generator written in the $\eta - {\cal P}$
order [i.e., all ghost momenta are put to the right of their conjugate
ghost variables by using repeatedly the ghost (anti)commutation relations]:
\begin{eqnarray}
\hat\Omega={\hat\eta}^o&&({1\over 2}  f^{-{1\over 2}} \hat p_i  g^{ij} f
\hat p_j  f^{-{1\over 2}} + {i\over 2} f^{1\over 2} c_{oa}^{aj}\hat p_j
f^{-{1\over 2}}+ \lambda) + \hat \eta^a f^{1\over 2} \xi^i_a \hat p_i
f^{-{1\over 2}}+\nonumber\\ &&+ {1\over 2} \hat \eta^o \hat \eta^a
(f^{1\over 2} c_{oa}^{bj}\hat p_j f^{-{1\over 2}} + f^{-{1\over 2}} \hat
p_j c_{oa}^{bj} f^{1\over 2} ) \hat {\cal P}_b + {1\over 2} \hat \eta^a
\hat \eta^b C_{ab}^c \hat {\cal P}_c.
\label{omeord'}\end{eqnarray} 
This definition has the following nice properties:\\ (i) In the limit
$\hbar \rightarrow 0$ they go over into the original classical
constraints;\\ (ii) they satisfy the first class conditions.

In this case, the coefficient of ${\hat\eta}^o$ in the Eq.  (\ref{omeord'})
is the quadratic constraint operator $\hat h$ and the coefficients of
${\hat\eta}^a$ are the supermomentum constraint operators ${\hat G}_a$.

\bigskip

So far we have dealt with a constant potential.  The introduction of a 
nonvanishing potential $\lambda \vartheta(q)$ in the BRST formalism
can be accomplished in a very simple way: by performing a unitary
transformation
\begin{equation}
\hat \Omega  \rightarrow  e^{i\hat  C}~\hat\Omega~ e^{-i\hat C},\label{tu}
\end{equation}
leading to a new Hermitian and nilpotent BRST generator. So let us choose
\begin{equation}\hat C={1\over 2}[\hat\eta^o~ ln~  \vartheta(q)~\hat{\cal
P}_o-\hat{\cal 
P}_o~ ln~ \vartheta(q)~\hat\eta^o],~~~~~~~\vartheta(q)>0.\label{de}
\end{equation}  
Thus,
\begin{eqnarray}\hat\Omega=&&{\hat\eta}^o({1\over 2}{\vartheta}^{1\over 2}
f^{-{1\over 2}}\hat p_i g^{ij}f \hat p_j f^{-{1\over 2}}{\vartheta}^{1\over
2}+ {i\over 2}{\vartheta}^{1\over 2}  f^{1\over 2} c_{oa}^{aj}\hat p_j
f^{-{1\over 2}} {\vartheta}^{1\over 2} +\lambda {\vartheta}) + \hat\eta^a
{\vartheta}^{-{1\over 2}}f^{1\over 2} \xi^i_a \hat p_i f^{-{1\over
2}}{\vartheta}^{1\over 2}\nonumber\\ &&+\hat\eta^o\hat\eta^a \xi^i_a (ln
{\vartheta})_{,i}\hat {\cal P}_o + {1\over 2} \hat \eta^o \hat \eta^a
{\vartheta}^{{1\over 2}} (f^{1\over 2} c_{oa}^{bj} \hat p_j f^{-{1\over
2}} + f^{-{1\over 2}} \hat p_j c_{oa}^{bj} f^{1\over 2}
){\vartheta}^{1\over 2} \hat {\cal P}_b + {1\over 2} \hat \eta^a \hat
\eta^b C_{ab}^c \hat {\cal P}_c.
\label{omegaesc}\end{eqnarray} 

The resulting operator $\hat \Omega$ corresponds to a quadratic constraint
$H\  =\  \vartheta\  h$ (then, $C^{bj}_{oa}\ =\ \vartheta\ c^{bj}_{oa}$).   The
constraint operators can be read in Eq. (\ref{omegaesc}):
\begin{equation}\hat H={1\over 2}{\vartheta}^{1\over 2}
f^{-{1\over 2}}\hat p_i g^{ij}f \hat p_j f^{-{1\over 2}}{\vartheta}^{1\over
2}+ {i\over 2}{\vartheta}^{-{1\over 2}}  f^{1\over 2} C_{oa}^{aj}\hat p_j
f^{-{1\over 2}} {\vartheta}^{1\over 2} +\lambda {\vartheta}
,\label{vcuadq'v}\end{equation}
\begin{equation}\hat G_a={\vartheta}^{-{1\over 2}}f^{1\over 2}
\xi^i_a\hat p_i f^{-{1\over 2}}{\vartheta}^{1\over 2},
\label{vlinq'}\end{equation} 
with the corresponding set of structure functions,
\begin{equation}\hat C_{oa}^o=\xi^i_a (ln ~{\vartheta})_{,i},\end{equation}
\begin{equation}\hat  C_{oa}^b = {1\over 2} \left(  {\vartheta}^{-{1\over  2}} 
f^{1\over  2}
C_{oa}^{bj}  \hat  p_j  f^{-{1\over  2}}  \vartheta^{1\over  2}  + 
\vartheta^{1\over 2} f^{-{1\over 2}} \hat p_j C_{oa}^{bj}
f^{1\over 2}  {\vartheta}^{-{1\over 2}} \right),
\end{equation}
\begin{equation}\hat C_{ab}^c=C_{ab}^c,\end{equation} 
all of them properly ordered for satisfying of the constraint algebra,
\begin{equation}[\hat  H,\hat  G_a]=\hat C_{oa}^o \hat H  +  \hat
C_{oa}^b(q,p) \hat G_b, \label{ccq'}\end{equation} 
\begin{equation}[\hat G_a,\hat G_b]=\hat C_{ab}^c(q) \hat G_c.
\label{clq'}\end{equation}

\section{Conclusions}

The  result  (\ref{vcuadq'v})  says  that  the  operator  associated  with  a
first  class    constraint    $H={1\over   2}  G^{ij}(q)p_ip_j+  V(q)$,  with
$V(q)>0~~\forall q$, is not the Laplacian for the metric
$G^{ij}$ plus $V$, but
\begin{equation}
\hat H={1\over 2} V^{1\over 2} f^{-{1\over 2}}
\hat  p_i  V^{-1} G^{ij} f \hat  p_j f^{-{1\over 2}}
V^{1\over 2}+{i\over 2} V^{-{1\over 2}}  f^{1\over 2} C_{oa}^{aj}\hat p_j
f^{-{1\over 2}} V^{1\over 2}    +V,\label{vcuadG}\end{equation} 
since the metric in the kinetic term must be $g_{ij}=V G_{ij}$.
For the sake of simplicity, we use a definite positive potential, but it
should be noted that, in  general,  what  is  required  is a nonvanishing one.

As it is well known, the BRST formalism provides  ghost  contributions to the
constraint  operators,  which  are needed for the satisfying of the  algebra
and/or  for  preserving  the  geometrical character of the wave function.   The
ghost  contribution  to  the  quadratic  constraint  is  the  second  term in
Eq. (\ref{vcuadG}) that  will be analyzed below.  The linear constraints acquire
two  anti-Hermitian terms  associated  with  the  traces  of  the  structure
functions:
\begin{equation}
\hat G_a= V^{-{1\over 2}} f^{1\over 2} \xi^i_a \hat p_i f^{-{1\over
2}} V^{1\over 2}=\xi^i_a \hat p_i - {i\over 2} \xi^i_{a,i} + {i\over 2}
C^b_{ab} + {i\over 2} C^o_{ao},\label{varlinqe}\end{equation}
where ${i\over 2} C^o_{ao} = -{i\over 2} \xi^i_a (\ln V)_{,i}$ is the
``cocycle'' of Ref.\ \onlinecite{hk90}.\footnote{Actually 
the restriction on the potential can be relaxed, because the results of 
Sec. III and IV do not change if $\lambda$, instead of being constant,
is a function
$\lambda(y)$ invariant on the gauge orbits associated with the linear 
constraints. The potential should be only restricted to factorizing
as $V = \vartheta(q) \lambda(y)$, $\vartheta(q) > 0$. This factorization
allows the existence of a globally well-defined ``physical gauge" in 
Ref. \cite{hk90}:
$\Omega (q)$ therein could be taken to be $\ln \vartheta(q)$. However, 
nonpositive definite potentials make less evident the way to build the
inner product in the physical Hilbert space\cite{bf}.}
Then, the kinetic term in the super-Hamiltonian and the supermomenta are
sensible to the existence of a potential.

\bigskip
The Hamiltonian constraint operator (\ref{vcuadG})  differs from the
one employed in Ref.\ \onlinecite{hk90}, where a curvature term was
introduced to retain the invariance of the theory under scaling.  Instead,
the invariance under scaling is provided by the role played by the
potential in the constraint operators.
In fact, the role played by the factors $f^{{\pm}{1\over 2}}$, $V^{{\pm}{1\over
2}}$ is clear whenever one pays attention to the transformations which
should leave invariant the theory; these are (i) coordinate changes, (ii)
combinations of the supermomenta [Eq.  (\ref{cbase})], and (iii)
scaling of the super-Hamiltonian ($H \rightarrow e^{\Theta}~H$).  The
physical gauge-invariant inner product of the Dirac wave functions,
\begin{equation}
(\varphi_1,\varphi_2)=\int dq\ \big[\prod^{m+1} \delta(\chi)\big]\ J\
\varphi^*_1(q)\ \varphi_2(q),\label{prodesc}\end{equation}
(where $J$ is the Faddeev-Popov determinant and $\chi$ are the $m+1$ gauge
conditions) must be invariant under any of these transformations.  On 
account of the change of the Faddeev-Popov determinant
under (ii) and (iii), the inner product will remain invariant if the Dirac
wave function changes according to\cite{fhp}
\begin{equation}
\varphi  \rightarrow    \varphi'=(det    A)^{1\over    2}
e^{-{\Theta\over 2}} \varphi.
\end{equation}
So, the factors $f^{{\pm}{1\over 2}}$, $V^{{\pm}{1\over 2}}$ in the constraint
operators  are just what are
needed in order that $\hat G_a\varphi,~\hat H\varphi,$ and ${\hat
C}_{oa}^b\varphi$ transform as $\varphi$, so preserving the geometrical
character of the Dirac wave function. 

Whenever the reader prefers to regard the wave function as invariant under
the relevant transformations (i)-(iii), he/she should perform the
transformation 

\begin{eqnarray}
\varphi ~&& \rightarrow   \phi=f^{-{1\over  2}}V^{+{1\over
2}}\varphi, \label{fun} \\ \hat{\cal O} ~ && \rightarrow f^{-{1\over
2}}V^{+{1\over 2}}\hat{\cal O} f^{+{1\over 2}}V^{-{1\over 2}}. 
\label{op} \end{eqnarray}
The corresponding physical inner product results in the integration of the
invariant $\phi^*_1\phi_2$ in the invariant volume $V^{-1}~J~[\prod
\delta(\chi)]~{\tilde\alpha}$.
\begin{equation}
(\phi_1,\phi_2)=(\varphi_1,\varphi_2)=\int \tilde\alpha\ V^{-1} \ J\
\big[\prod \delta(\chi)\big]\ \phi^*_1\ \phi_2.\label{peinv}\end{equation}

\bigskip
Since the inner product (\ref{prodesc}) or (\ref{peinv}) is
invariant under the transformation (ii), one can choose the Abelian
coordinate basis $\vec\xi_a^{\prime}={\partial/ \partial Q^a}$
($G^{\prime}_a=P_a$). Thus, the volume reads 
\begin{equation}
{\tilde\alpha}^{\prime}=dQ^1 \wedge ...   \wedge  dQ^m  \wedge \omega(y) dy^1
\wedge ... dy^{n-m}.\label{volab}
\end{equation}

 Then, the linear constraint equations for the invariant
Dirac wave function $\phi$ are
\begin{equation}
{\partial\phi\over\partial Q^a} = 0.\label{linfacil}
\end{equation}
These  equations notably simplify the super-Hamiltonian  constraint  
equation which, when written in the coordinate basis $\{dQ^a, dy^r\}$ [then 
$f' = \omega(y)$], reduces to  
\begin{equation}
\left(-{1\over  2}  V  {\partial\over\partial  Q^a}\ V^{-1}  G^{ar}
{\partial\over\partial y^r}
-{1\over 2} V{\omega(y)}^{-1}{\partial\over\partial y^r}\ {\omega(y)}V^{-1} 
G^{rs} {\partial\over\partial y^s}
+ {1\over 2} C^{ar}_{oa} {\partial\over \partial y^r} + V\right)\ \phi    =
0.\label{casilaplace}
\end{equation}
The potential can be factorized out. Then, taking into account the Eqs.
(\ref{cc})  and  (\ref{fec}),  it  is    $V^{-1}C^{br}_{oa}= c^{br}_{oa}  
=  \partial g^{br}/\partial Q^a$ and $V^{-1}G^{rs}=g^{rs} = g^{rs}(y)$. Thus, 
\begin{equation}
\left(-{1\over    2}  \omega(y)^{-1}  {\partial\over\partial  y^r}\ \omega(y)
g^{rs}(y) {\partial\over\partial y^s} + 1\right)\ \phi =  0.\label{laplac}
\end{equation}
Therefore,  the ghost contribution to the  super-Hamiltonian  allows  for  the
emerging of a ``Laplacian" in terms of  the  reduced variables $\{y^r\}$.  In
order to obtain the true Laplacian for the scale-invariant reduced metric
$g_{rs}(y)$,    one    should    choose    $\omega(y)$   to  be  $\vert  \det
(g_{rs})\vert^{1/2}$. It is clear that the BRST formalism cannot give a value
for $\omega(y)$,  because  $\hat\Omega$  is  Hermitian and nilpotent whatever
$\omega(y)$ is.\footnote{If, for instance, the system only had a quadratic
constraint $h$, then $\hat\Omega  =  \hat\eta  \hat h$ would be Hermitian and
nilpotent for any Hermitian ordering of $\hat h$.}

\bigskip

In order to glance at the relationship between Dirac quantization and
reduced space quantization, let us choose the gauge-fixing functions
$\chi^a=Q^a,~ \{\chi^o,P_a\}=0$. 
Then, one integrates the $Q^a$'s in Eq. (\ref{peinv}) using the volume 
$\tilde{\alpha}^{\prime}$ to obtain
\begin{equation}
(\phi_1,\phi_2)=\int \tilde\omega\ V^{-1} \ J_o\ \delta(\chi^o)\
\phi^*_1[q^i(Q^a=0,y^r)]\ \phi_2[q^i(Q^a=0,y^r)].\label{pered}\end{equation}

One can define a density of weight 1/2 under changes of the $y^r$'s:
\footnote{The function $\omega^{-1} f'$ plays
the role of $\mu$ in Ref. \cite{fhp} and of ${\cal M}$ in Ref. \cite{bar} . In
fact, let us use the Abelianized basis ${\vec\xi}^{\prime}_a$ in $T_{||}M$
which is a coordinate basis: ${\vec\xi}^{\prime}_a={\partial/ \partial
Q^a}$ ($G^{\prime}_a=P_a$), ${\tilde E}^{\prime a}=dQ^a$.  Then, 
$\omega^{-1} f'$  is the  Jacobian  of  the  coordinate  change  $(Q^a,y^r)
\rightarrow q^i$.}

\begin{eqnarray}
\varphi_R(y)&&=\omega (y)^{1\over  2}\   V[q^i(Q^a=0,y^r)]^{-{1\over 2}}\
\phi[q^i(Q^a=0,y^r)]\nonumber\\ &&=\omega (y)^{1\over 2}\ f'
[q^i(Q^a=0,y^r)]^{-{1\over 2}}\ \varphi[q^i(Q^a=0,y^r)].\label{densred}
\end{eqnarray} 

Then, 
\begin{equation}
(\phi_1,\phi_2)=(\varphi_{R_1},\varphi_{R_2})=\int dy \ J_o\ \delta(\chi^o)
\varphi^*_{R_1}(y) \varphi_{R_2}(y),\label{peredens}\end{equation}

and $\varphi_R$ is the Dirac  wave  function in a ``reduced" space where only
the quadratic constraint remains:  $\varphi_R$ is  constrained  by  the Eq.
(\ref{laplac}) satisfied by $\phi [q^i(Q^a=0,y^r)]$.

\bigskip

We close the conclusions by giving a beautiful classical argument
supporting the 
inclusion of the potential in the kinetic term. Generally covariant systems
are invariant under changes of the parameter in the functional
action\cite{ht}. This means that the parameter is physically irrelevant: it
is not the time. The time could be hidden among the dynamical variables
and, as a result, the Hamiltonian is constrained to vanish\cite{kuc}. In this
case the time would be identified as a function $t(q,p)$ in phase space
that monotonically increases on all the dynamical trajectories. 
Since  the  here-studied $H$ is equivalent  to  a  super-Hamiltonian  with  a
constant potential,
the systems embraced in this article are those resembling a relativistic
particle in a curved spacetime.
Then, the time is hidden in the configuration
space ({\it intrinsic time}\cite{kuc}). This means that the trajectory in
the configuration space contains all the dynamical information about the
system. In classical mechanics, the Jacobi's principle\cite{lanczos} is the
variational principle for getting the paths in configuration space, for a
fixed energy $E$, without information about the evolution of the system in
the parameter of the functional action. The paths are obtained by varying
the functional
\begin{equation}
I=\int_{q'}^{q''} \sqrt {2 |E-V|G_{ij}dq^i dq^j}.\label{jac}
\end{equation}

In our case, the energy is zero, and Eq. (\ref{jac}) looks such as the
functional 
action of a relativistic particle of mass unity in a curved background with
metric $2 V G_{ij}=2g_{ij}$. The paths are geodesics of this
metric $2 g_{ij}$, instead of $G_{ij}$.  When the gauge is fixed to be $Q^a =
0$, the Jacobi's principle reduces to the variation of the functional
\begin{equation}
I_R = \int_{y'}^{y''} \sqrt {2 V G_{rs} dy^r dy^s} ,\label{jac2}
\end{equation}
so  giving the classical support to the constraint  equation  (\ref{laplac}),
where the Laplacian is the one associated with the scale-invariant metric
$2 V G_{rs} = 2 g_{rs}(y)$ appearing in Eq. (\ref{jac2}).

\acknowledgments 

We would like to thank Marc Henneaux for helpful discussions.  This
research was supported by Universidad de Buenos Aires (Proy. EX 252), 
Consejo Nacional
de Investigaciones Cient\'\i ficas y T\'ecnicas and Fundaci\'on Antorchas.

\end{document}